\documentclass[11pt,twoside]{article}
\usepackage{asp2010}

\resetcounters

\begin{document}

\title{CANFAR+Skytree: A Cloud Computing and Data Mining System for Astronomy}
\author{Nicholas M. Ball$^1$
\affil{$^1$National Research Council Canada, 5071 West Saanich Road, Victoria, BC V9E 2E7}}

\begin{abstract}
To-date, computing systems have allowed either sophisticated analysis of small datasets, as exemplified by most astronomy software, or simple analysis of large datasets, such as database queries. At the Canadian Astronomy Data Centre, we have combined our cloud computing system, the Canadian Advanced Network for Astronomical Research (CANFAR), with the world's most advanced machine learning software, Skytree, to create the world's first cloud computing system for data mining in astronomy.

CANFAR provides a generic environment for the storage and processing of large datasets, removing the requirement for an individual or project to set up and maintain a computing system when implementing an extensive undertaking such as a survey pipeline. 500 processor cores and several hundred terabytes of persistent storage are currently available to users, and both the storage and processing infrastructure are expandable. The storage is implemented via the International Virtual Observatory Alliance's VOSpace protocol, and is available as a mounted filesystem accessible both interactively, and to all processing jobs. The user interacts with CANFAR by utilizing virtual machines, which appear to them as equivalent to a desktop. Each machine is replicated as desired to perform large-scale parallel processing. Such an arrangement enables the user to immediately install and run the same astronomy code that they already utilize, in the same way as on a desktop. In addition, unlike many cloud systems, batch job scheduling is handled for the user on multiple virtual machines by the Condor job queueing system.

Skytree is installed and run just as any other software on the system, and thus acts as a library of command line data mining functions that can be integrated into one's wider analysis. Thus we have created a generic environment for large-scale analysis by data mining, in the same way that CANFAR itself has done for storage and processing. Because Skytree scales to large data in linear runtime, this allows the full sophistication of the huge fields of data mining and machine learning to be applied to the hundreds of millions of objects that make up current large datasets.

We demonstrate the utility of the CANFAR+Skytree system by showing science results obtained, including assigning photometric redshifts to the MegaPipe reductions of the Canada-France-Hawaii Telescope Legacy Wide and Deep surveys. This project involves producing, handling, and running data mining on, a catalog of over 13 billion object instances. This is comparable in size to those expected from next-generation surveys, such as the Large Synoptic Survey Telescope.

The CANFAR+Skytree system is open for use by any interested member of the astronomical community.
\end{abstract}

\section{CANFAR} \label{Sec: CANFAR}

The Canadian Advanced Network for Astronomical Research (CANFAR;\footnote{\url{http://canfar.phys.uvic.ca}} \citeauthor{gaudet:canfar} \citeyear{gaudet:canfar}) is the cloud computing system of the Canadian Astronomy Data Centre (CADC). It is the first system designed to provide this capability to astronomers. It provides 500 processor cores, up to 32G memory per processor node, and several hundred terabytes of storage, implemented via the International Virtual Observatory Alliance-compliant VOSpace system. VOSpace can be accessed both interactively and in batch as a mounted filesystem, via VOFS\footnote{\url{http://canfar.astrosci.ca/wiki/index.php/VOSpace_filesystem}}. The user interacts with the system via a Virtual Machine, and submits batch processing jobs via Condor. The upshot is that one can configure and run one's own code as on a desktop, then replicate it across the cloud. Other science clouds exist, such as the LHC Grid\footnote{\url{http://wlcg.web.cern.ch}}, and the EU Helix Nebula project\footnote{\url{http://www.helix-nebula.eu}}, but none provides the facilities for astronomers of CANFAR, or the analytics capability of Skytree.

\section{Skytree} \label{Sec: Skytree}

\begin{figure}
\plotfiddle{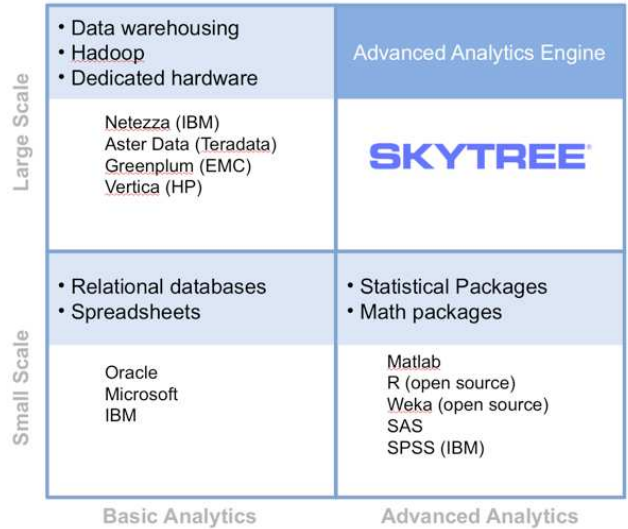}{7cm}{0}{40}{40}{-125}{0}
\caption{Skytree allows both advanced analytics, and its application to large data. This fills a vacant parameter space that is essential for future astronomical data analysis. \label{Fig: Skytree}}
\end{figure}

Skytree is the world's most advanced machine learning software. It acts as a machine learning server to allow advanced data mining on large data (Figure \ref{Fig: Skytree}), e.g., within one's data processing pipeline, or more specialized science project. Skytree's A. Gray also heads the FASTlab\footnote{\url{http://www.fast-lab.org}} group at the Georgia Institute of Technology. The group holds several records for the fastest implementation of well-known machine learning algorithms. Algorithms that otherwise scale as, e.g., $N^2$, for $N$ objects, are implemented to scale linearly, without loss of accuracy. While each specific use case will remain science-driven, the underlying tools are not dataset-specific. Thus, the installation of Skytree on the CADC infrastructure makes possible the practical use of these algorithms by astronomers who are not data mining specialists. The software's quality and robustness renders it suitable for publication-quality research.

\section{Performance} \label{Sec: Performance}

\begin{figure}
\plotfiddle{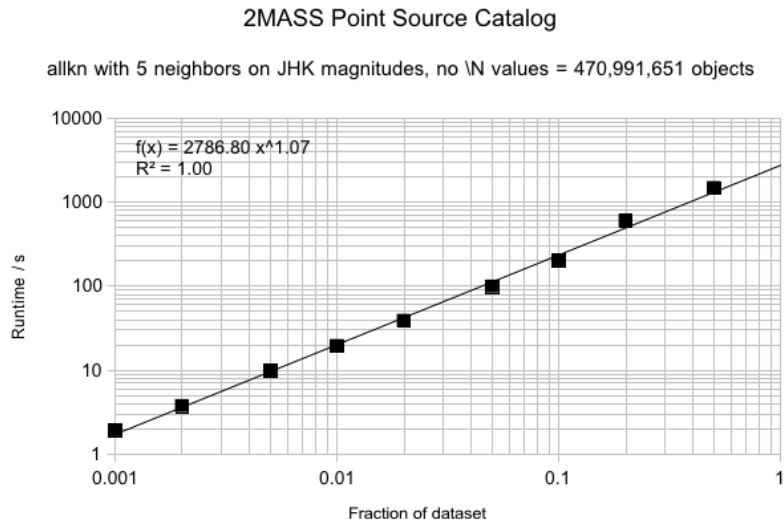}{7cm}{0}{50}{50}{-155}{0}
\caption{Typical performance benchmarks. Linear runtime scaling with dataset size: Skytree's {\tt allkn} has computed the 5 nearest neighbors of each dataset point. A na\"{\i}ve implementation would be quadratic, and the best fit line would have a slope of order 2, not 1. Runtime error bars are less than the size of the points on the plot, with the exception of the 0.2 and 0.5 fractions, for which they are not yet computed. These two utilize CANFAR nodes with 256G memory. \label{Fig: runtime}}
\end{figure}

Skytree claims to make fundamental machine learning algorithms scalable. We have tested these claims on the CANFAR+Skytree system, and have verified that it scales to the largest current surveys, and beyond to upcoming datasets. To do this, we analyze large astronomical catalogs. An example benchmark scaling is shown in Figure \ref{Fig: runtime}.

\section{Science Example} \label{Sec: Science Example}

We have computed photometric redshifts for Canada-France-Hawaii Telescope Legacy Survey (CFHTLS). Most studies use single values, or a Gaussian approximation. This is in general unsuitable, and full probability density functions (PDFs) in redshift produce superior results \citep{ball:pdfphotoz}. Skytree's {\tt allkn} and {\tt kde} allow us to produce these PDFs for the CFHTLS nonparametrically, utilizing the full information within the training set. This involves handling a catalog of 13 billion objects, comparable in size to upcoming Large Synoptic Survey Telescope project.

\section{Conclusions}

CANFAR+Skytree represents world's first cloud computing system for data mining in astronomy, and is open for use by any interested member of the astronomical community. For further details on usage, see the ADASS XXII Focus Demo (\citeauthor{ball:adass12focusdemo} \citeyear{ball:adass12focusdemo}, this volume), or visit the CANFAR+Skytree website at \url{https://sites.google.com/site/nickballastronomer}.

\acknowledgements This research used the facilities of the Canadian Astronomy Data Centre, operated by the National Research Council of Canada with the support of the Canadian Space Agency. Funding for CANFAR was provided by CANARIE via the Network Enabled Platforms Supporting Virtual Organisations program. The author thanks D. Schade, A. Gray and M. Hack for their contributions to this work.

\bibliography{P062_refs}

\end{document}